\documentclass[
reprint,
superscriptaddress,
nofootinbib,
nobibnotes,
amsmath,amssymb,
aps,
pre,
floatfix,
]{revtex4-2}

\usepackage{dcolumn}
\usepackage{bm}
\usepackage{graphicx}
\graphicspath{{figures/}}
\usepackage{amsmath}
\usepackage{algorithm}
\usepackage{booktabs}
\usepackage{algpseudocode}
\usepackage{natbib}
\usepackage{layouts}
\usepackage[colorlinks=true,linkcolor=blue, citecolor=blue, urlcolor=blue]{hyperref}
\usepackage[capitalize,noabbrev]{cleveref}
\begin{document}
\preprint{APS/PRE}

\newcommand{\mat}[1]{\mathbf{#1}} 
\renewcommand{\vec}[1]{\bm{\mathrm{#1}}} 

\title{Data-driven Modeling of Granular Chains with Modern Koopman Theory}

\author{Atoosa Parsa}
\email{atoosa.parsa@tufts.edu}
\affiliation{Department of Biology, Tufts University, Medford, Massachusetts, USA}

\author{James Bagrow}
\email{james.bagrow@uvm.edu}
\affiliation{Department of Computer Science, University of Vermont, Burlington, Vermont, USA}

\author{Corey S. O'Hern}
\email{corey.ohern@yale.edu}
\affiliation{Department of Mechanical Engineering and Materials Science, Yale University, New Haven, Connecticut, USA}

\author{Rebecca Kramer-Bottiglio}
\email{rebecca.kramer@yale.edu}
\affiliation{Department of Mechanical Engineering and Materials Science, Yale University, New Haven, Connecticut, USA}

\author{Josh Bongard}
\email{josh.bongard@uvm.edu}
\affiliation{Department of Computer Science, University of Vermont, Burlington, Vermont, USA}


\begin{abstract}
Externally driven dense packings of particles can exhibit nonlinear wave phenomena that are not described by effective medium theory or linearized approximate models. Such nontrivial wave responses can be exploited to design sound-focusing/scrambling devices, acoustic filters, and analog computational units. At high amplitude vibrations or low confinement pressures, the effect of nonlinear particle contacts becomes increasingly noticeable, and the interplay of nonlinearity, disorder, and discreteness in the system gives rise to remarkable properties, particularly useful in designing structures with exotic properties. 
In this paper, we build upon the data-driven methods in dynamical system analysis and show that the Koopman spectral theory can be applied to granular crystals, enabling their phase space analysis beyond the linearizable regime and without recourse to any approximations considered in the previous works. We show that a deep neural network can map the dynamics to a latent space where the essential nonlinearity of the granular system unfolds into a high-dimensional linear space. As a proof of concept, we use data from numerical simulations of a two-particle system and evaluate the accuracy of the trajectory predictions under various initial conditions. By incorporating data from experimental measurements, our proposed framework can directly capture the underlying dynamics without imposing any assumptions about the physics model. Spectral analysis of the trained surrogate system can help bridge the gap between the simulation results and the physical realization of granular crystals and facilitate the inverse design of materials with desired behaviors.

\end{abstract}


\maketitle

\section{Introduction}
\label{sec:introduction}
Granular materials are athermal ensembles of macroscopic noncohesive particles in which adjacent particles interact with elastic repulsive forces only when they come into contact. Such particulate systems are unique in that depending on the amount of confining pressure and driving forces, they can exhibit characteristics of any of the three states of matter, namely solid, liquid, and gas. Here, we study confined 1D granular systems in a jammed state, where the packing fraction determines the strength of the (extrinsic) nonlinearity in the system.

Granular crystals have gained extensive attention from a diverse range of disciplines in the last decade, resulting in a burst of progress, both in their theoretical foundations and experimental realizations \cite{Chong.etal2016}. They are utilized as architectured structures in applications, including energy localization and vibration absorption layers \cite{Zhang.etal2015, Taghizadeh.etal2021b}, acoustic computational units like switches and logic elements \cite{Li.etal2014a, Parsa.etal2023}, granular actuators \cite{Eristoff.etal2022}, grippers \cite{Brown.etal2010}, acoustic filters \cite{Boechler.etal2011, Lawney.Luding2014}, and sound focusing/scrambling devices \cite{Porter.etal2015, Spadoni.Daraio2010}. Beyond practical applications, granular assemblies are also studied as simplified test beds for investigating fundamental problems in many disciplines, including materials science and condensed matter physics and providing insights to understand other complex nonequilibrium systems \cite{Jaeger.Nagel1992, Jaeger.etal2000a}.

\subsection{Related Work}
\label{sec:relatedWorks}
Finding coherent spatial and temporal structures in the phase space of driven high-dimensional granular crystals has been an active area of research in the past decades \cite{Dove1993, Friesecke.Wattis1994, Narisetti2010}. Despite the long history of scientific research devoted to the study of granular materials, there are still no general methodologies for analyzing the relation between their constituents' material and geometric properties and the emergent mechanical responses. The granular crystal is a nonlinear, non-integrable, and high-dimensional system, and explicitly solving the differential equations of motion to find the exact solutions is usually not feasible. Therefore, most studies rely on numerical simulations and computational techniques with simplifying assumptions such as linearity, periodicity, and state-space continuity.
\begin{figure*}[htp!]
\centering
\includegraphics[width=1.0\textwidth]{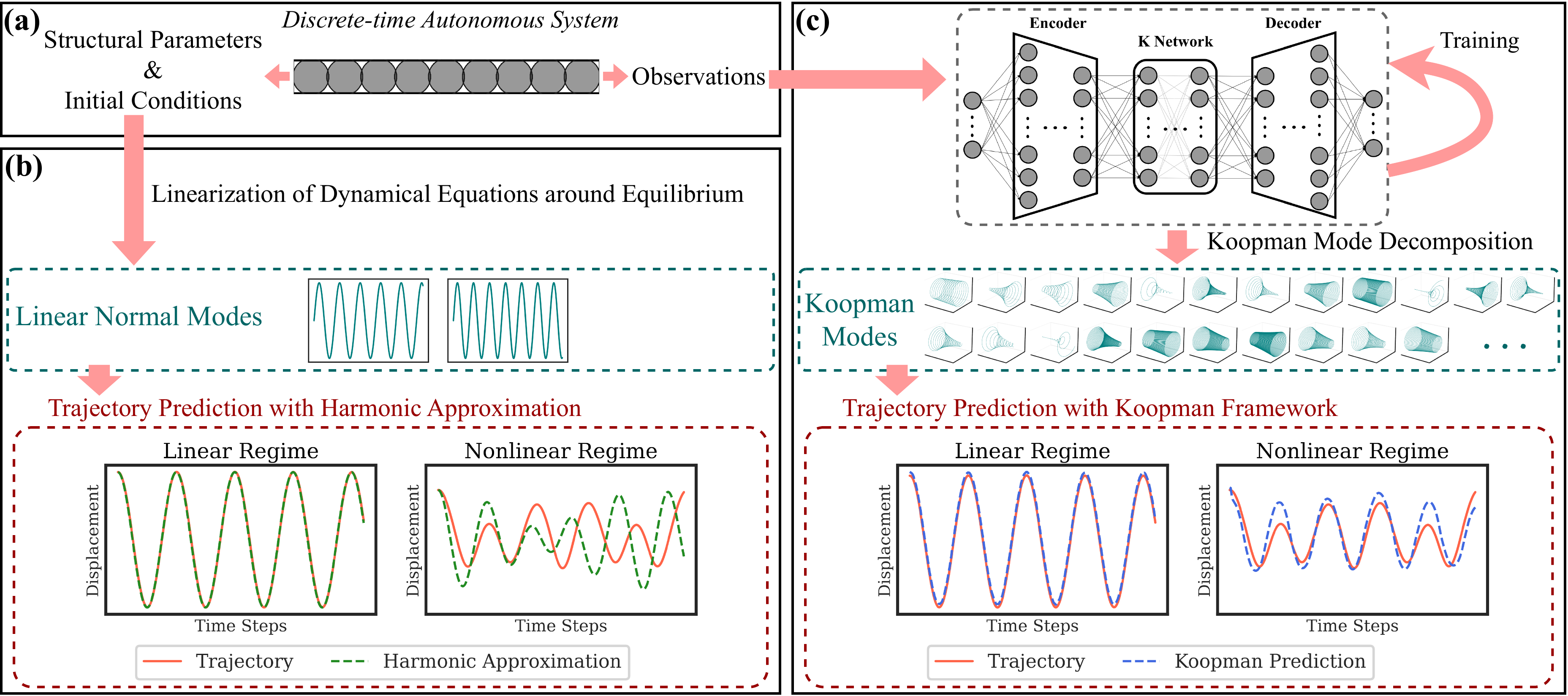}
\caption{\label{fig:overview}
Overview of the proposed method. (a) A chain of elastic spherical particles is studied as a discrete-time autonomous system. Given the structural parameters (i.e. particles' size, mass, and stiffness) and initial conditions (positions and velocities of the particles), the system can be numerically simulated to obtain the trajectories of particle displacements in time. (b) The harmonic balance method predicts the particle trajectories as a superposition of the Linear Normal Modes (LNMs). Such prediction is a good approximation only in the linear regime. (c) Our proposed data-driven framework can predict particle trajectories in linear and nonlinear regimes by training a deep neural network using real or simulated system observations. Instead of LNMs, an unlimited number of complex-valued Koopman modes are obtained from the trained \textit{K Network} to provide a more accurate prediction of particles' trajectories in the strongly nonlinear regime.}
\vskip -0.1in
\end{figure*}

Most computational techniques in granular systems rely on linearizing the equations of motion around the equilibrium state to find a closed-form dispersion relation. Dispersion relation provides valuable insight into the time-periodic wave solutions of the unforced lattice systems and is a central topic in the spectral analysis of lattice dynamics \cite{Dove1993}. Similar analyses have been adopted in the granular crystals in the continuum limit \cite{Friesecke.Wattis1994}. 

For example, in a strongly compressed granular chain, the equations of motion can be simplified (when Taylor expanded) to those of a spring-mass system with nonlinear springs, resembling the well-studied Fermi-Pasta-Ulam (FPU) lattice with polynomial potential \cite{Ford1992}. However, there are interesting nonlinear wave phenomena that are a direct consequence of the discrete nature of the system and will not be captured with such continuum approximations because the contributing spatial scales are at the particle contact scale and below the scale of lattice spacing \cite{Nesterenko.etal2005}.

For the infinite chain, the analytical dispersion relation can be found similarly to methods used for weakly nonlinear lattices, where the solution is expressed as a perturbation from the linear case. In the short wavelength limit, perturbation techniques such as the Lindstedt-Poincar\'e method and the method of multiple scales have been used to obtain dispersion curves. For systems with strong nonlinearity, methods such as generalized harmonic balance can provide approximate wave solutions \cite{Narisetti2010}. Many of the aforementioned techniques are only applicable to homogeneous or periodically disordered systems because translational symmetry is essential for some mathematical constructs such as the Fourier series expansion that are used in the method. For systems with a periodic disorder, one needs to consider a reduced-order system by defining a custom unit cell and extending the analysis from the homogeneous case \cite{Starosvetsky.etal2012}. These methods also rely on the infinity assumption, meaning that they assume the waves will not reach the system's physical boundaries and therefore, find the non-reflecting plane wave solutions.  

The above analyses do not provide insights about the non-autonomous system, where we are interested in finding the periodic vibrational modes that are excited under various input driving frequencies and amplitudes. The dispersion relation in nonlinear systems is amplitude-dependent, and the passband and cut-off frequency change with the magnitude of the input vibrations. Due to the one-sided interparticle potentials, contact-breaking events can occur, and the system's behavior can diverge significantly from the continuum limit predictions. The dynamics can even become chaotic as the amplitude is increased beyond the weakly nonlinear regime. Most studies determine the variation of the band structure qualitatively by performing perturbation analysis as the wave amplitude varies. Stability analysis of the system's dynamic response has been investigated through numerical studies, where time-periodic solutions are found for various frequency and excitation amplitudes by methods such as Newton iterations \cite{Ganesh.Gonella2017}. The spectral stability of the solutions is then determined through the computation of Floquent multipliers corresponding to the periodic wave solution \cite{Boechler.etal2011}.

In conclusion, various dynamic phenomena in granular crystals such as solitary waves, discrete breathers, and dispersive shock waves need a complete consideration of their intrinsic characteristics including discreteness, disorder, and strong one-sided nonlinearity. However, the available methods for analyzing dynamic wave solutions usually rely on simplifying assumptions that ignore those characteristics to provide computationally tractable techniques. Therefore, a systematic study of spectral (wavenumber-frequency domain) and spatial (space-time domain) properties of granular crystals is still missing.

\subsection{Contributions}
\label{sec:contributions}
The interplay of discreteness, nonlinearity, and disorder can give rise to fascinating phenomena in the driven granular crystals. Acoustic metamaterials consisting of assemblies of particles with different material properties have tremendous potential to be used in real-life applications. Configurational and parametric freedom in constructing such artificial composite materials is essentially limitless. However, traversing such a high-dimensional non-intuitive design space is a daunting task, and developing effective methodologies capable of capturing their nonlinear dynamics is an open research problem.

In this paper, we propose using data-driven modal analysis methods to study the nonlinear wave propagation in granular crystals. We focus on elastic wave propagation in a Hamiltonian one-dimensional granular crystal (granular chain) and present a data-driven framework based on Koopman theory for modal analysis of strongly nonlinear systems. We show that the model can be trained using datasets made from measurements of the system's observables (in simulation or experiment).

\section{Granular Chains}
\label{sec:granularChains}
\subsection{Physics Model}
\label{sec:physicsModel}
\begin{figure}[htp!]
\centering
\includegraphics[width=1.0\columnwidth]{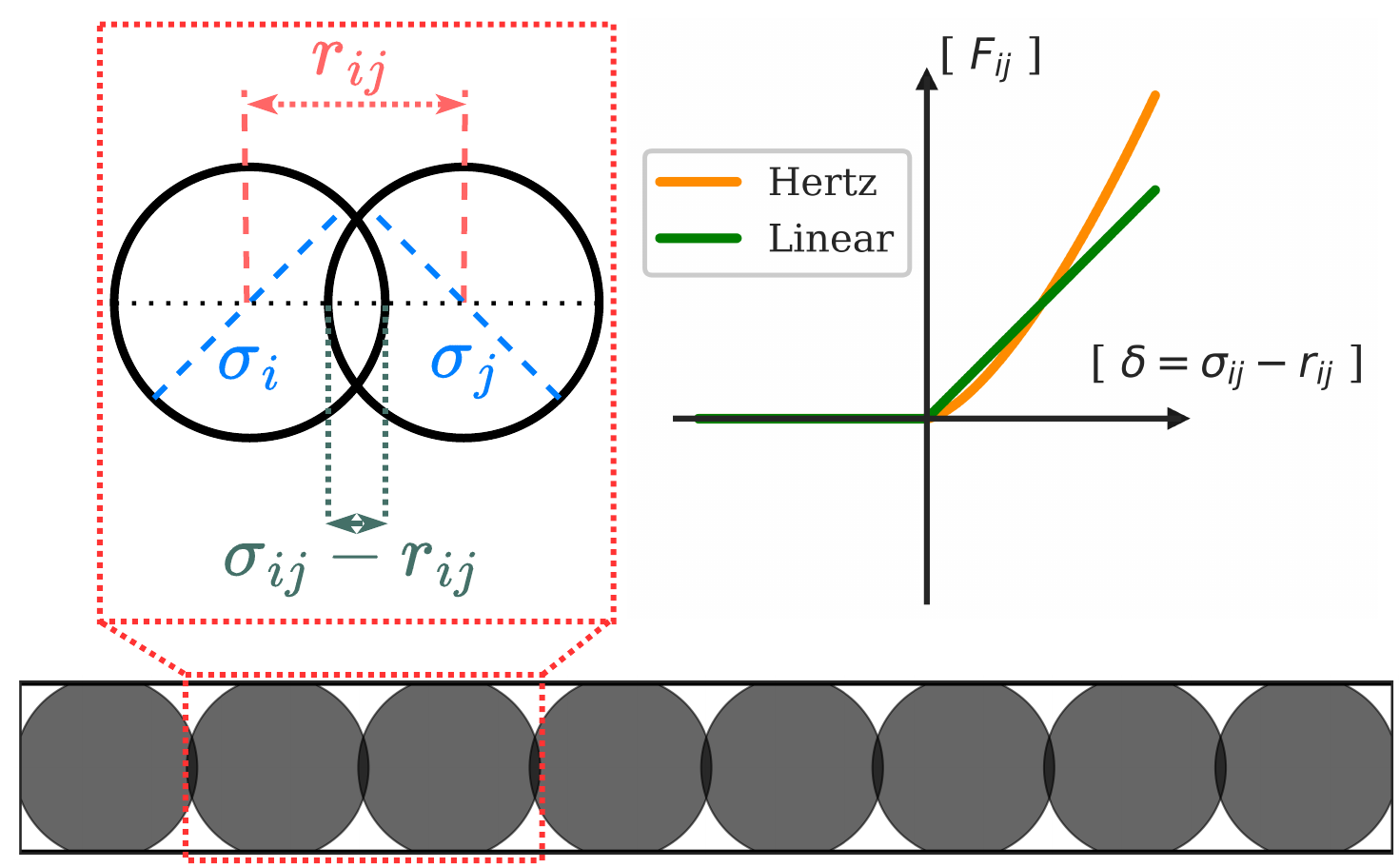}
\caption{\label{fig:setup} A granular chain that is made of two types of elastic spherical particles. \textit{Hertz's law} \cite{Hertz1882} describes the relation between the particle's overlap ($\delta=\sigma_{ij}-r_{ij}$) and applied force ($F$) as $F=\alpha \delta ^ \beta$. Here, $\beta$ is a constant that depends on the particle geometry and determines the nonlinearity of the contact forces. A commonly used value for spherical contacts is $\beta = \frac{3}{2}$, which produces a cubic nonlinearity in the equations of motion.
}
\end{figure}

The granular crystals discussed in this paper are finite-length one-dimensional configurations of spherical particles with identical diameters and variable elasticity placed on a horizontal flat surface (\cref{fig:setup}).

The system is a macroscopic scale granular system (particle sizes are in the millimeter-to-centimeter range), so the only forces acting on each particle are the finite-range repulsive interparticle contact forces. On the scale of particle contacts, we consider normal forces resulting from the adjacent particles' overlaps and ignore the tangential forces and particle rotations. 
With these assumptions, the local potential $V_{ij}$ between each pair of particles $i$ and $j$ can be written as
\begin{equation}
    \label{eq:potential}
    V_{ij}(r_{ij}) = \frac{\epsilon}{\alpha}{\left(1-\frac{r_{ij}}{\sigma_{ij}}\right)}^\alpha \; \Theta \left(1 - \frac{r_{ij}}{\sigma_{ij}}\right),
\end{equation}
where $\epsilon$ is the characteristic energy scale, $\alpha$ (discussed below) parameterizes the contact force nonlinearity, $r_{ij}$ is the particles' separation, and $\sigma_{ij}$ is the center-to-center separation at which the particles are in contact without any deformation ($\sigma_{ij}=\frac{\sigma_i+\sigma_j}{2}$ in the case of spherical particles with diameters $\sigma_i$ and $\sigma_i$), and $\Theta$ is the Heaviside function. 
The Heaviside function ensures that the potential field is one-sided, meaning that the particles only affect their adjacent neighbors when they overlap:
\begin{equation}
    \Theta \left(1 - \frac{r_{ij}}{\sigma_{ij}}\right) = 
        \left\{ \begin{aligned}
            & 0 & \quad r_{ij} \geq \sigma_{ij}, \\
            & 1 & \quad r_{ij} < \sigma_{ij}.
        \end{aligned} \right.
\end{equation}
The separation between two spherical particles is computed based on their Cartesian coordinates:
\begin{equation}
    |{r}_{ij}| = |\vec{r}_i-\vec{r}_j| = \sqrt{x_{ij}^2 + y_{ij}^2}.
\end{equation}

This is the simplest model for a granular chain that neglects special aspects such as particles' rotation and alignment, which might be more important in higher dimensional experimental setups but are negligible in smaller scales.

As mentioned, in \cref{eq:potential}, $\alpha$ determines the nonlinearity of the contact force. In this paper, we consider two cases: linear ($\alpha=2$) and Hertzian ($\alpha=\frac{5}{2}$). Interparticle forces $F_{ij}$ can be obtained by taking the derivative of the potential ($V_{ij}$) with respect to the displacement:
\begin{equation}
    \label{eq:force}
    \begin{aligned}
    F_{ij} & = -\frac{\partial V_{ij}(r_{ij})}{\partial r_{ij}} \\
    & = \frac{\epsilon}{\sigma_{ij}} \; {\Bigl(1-\frac{r_{ij}}{\sigma_{ij}}\Bigr)}^{\alpha-1} \; \Theta \Bigl(1 - \frac{r_{ij}}{\sigma_{ij}}\Bigr) \; \frac{\partial r_{ij}}{\partial (\text{$x_{ij}$ or $y_{ij}$})}.
    \end{aligned}
\end{equation}

We assume that the particles have equal mass ($m$) but can have different stiffness values. In this case, $\epsilon$ can be calculated using the effective stiffness:
\begin{equation}
    \epsilon_{ij} = 
    \left\{ \begin{aligned}
        k_i = k_j \quad  & \text{if $k_i=k_j$}, \\
        \frac{k_i k_j}{k_i + k_j} \quad  & \text{if $k_i \neq k_j$}. 
    \end{aligned}\right.
\end{equation}

It is worth mentioning that with this formulation, the equations of motion for the granular crystal are reminiscent of the mathematical model of a Fermi-Pasta-Ulam (FPU) oscillator~\cite{Ford1992}. 
Using the above notation, we can write Newton's equations of motion as
\begin{equation}
    \label{eq:motion}
    m \ddot{r}_{i} = F_{i} = \sum^N_{j = 1, j \neq i} F_{ij} + F_{ext},
\end{equation}
where the first term is the total force from the neighboring particles, and the second term is the system's external forces, which include the interaction force from the walls (in case of a fixed boundary condition) and the excitation applied to the system in the form of harmonic vibrations. 
Using \cref{eq:force}, we can obtain the partial forces in a one-dimensional system in $x$ direction:
\begin{equation}
        F^x(r_{ij}) = \frac{\epsilon_{ij}}{\sigma_{ij}} {\left(1-\frac{x_{ij}}{\sigma_{ij}}\right)}^{\alpha-1} \; \Theta \left(1-\frac{x_{ij}}{\sigma_{ij}}\right),
\end{equation}
\begin{equation}
        F^x_{iw} = \frac{\epsilon}{\sigma_{i}/2} \; {\left(1-\frac{x_{i}-x_{w}}{\sigma_{i}/2}\right)}^{\alpha-1} \; \Theta \left(1-\frac{x_{i}-x_{w}}{\sigma_{i}}\right),        
\end{equation}
where $F^x_{iw}$ is the force between particle $i$ and the wall placed at $x_w$.

\subsection{Harmonic Approximation}
\label{sec:harmonicApprox}
The harmonic approximation is the standard treatment for studying the oscillatory motion of many-body systems such as granular crystals. Despite its simplicity, this method provides an accurate prediction of future system states in the linear regime and is the basis of more advanced nonlinear methods of trajectory prediction and analysis. This section provides an overview of harmonic approximation in a granular chain described in \cref{sec:physicsModel}.

For a chain of $N$ particles, the total potential energy of the system ($U_t$) at time $t$ can be written in terms of the pairwise particle potentials (\cref{eq:potential}) as\footnote{The mathematical notation in this section has been adapted from \cite{Tanguy.etal2002, Tanguy.etal2010, Mizuno.etal2016, Xu2005, Schreck2012, Owens2012, Bertrand2016}.}
\begin{equation}
    U_{t}(\vec{r}) = \sum_{i=1}^{N-1} \sum_{j=i}^N V_{ij}(r_{ij}).
\end{equation}
As the particles start to vibrate from their equilibrium position ($\delta_{0, i}$), the difference in the total potential energy due to a displacement $\vec{u}$ in particle positions ($\vec{r}=\vec{\delta}_0+\vec{u}$) can be written as
\begin{equation}
    \Delta U = U_t(\vec{r}) - U_0(\vec{\delta}_0),
\end{equation}
where $U_0$ is the total energy at equilibrium on account of the static precompression. 
We can Taylor expand the above equation around the equilibrium, and, because $\nabla U_0 = 0$, we will have:
\begin{equation}
\begin{aligned}
    \Delta U &= U_t(\vec{r}) - U_0(\vec{\delta}_0) \\
    &= \frac{\partial U}{\partial \vec{r}}\Bigg|_{\vec{\delta}_0}\hspace{-0.5em}(\vec{r}-\vec{\delta}_0) + \frac{1}{2} \; (\vec{r}-\vec{\delta}_0) \; \frac{\partial^2 U}{\partial \vec{r}^2}\Bigg|_{\vec{\delta}_0}\hspace{-0.5em}(\vec{r}-\vec{\delta}_0) \\
     & \quad + \mathcal{O}(\vec{r}-\vec{\delta}_0)^3.
\end{aligned}
\end{equation}

Ignoring the higher-order ($\mathcal{O}(\vec{r}-\vec{\delta}_0)^3$) terms and noting that the net force at equilibrium is zero will give the energy difference in terms of the displacement ($\vec{u}$) and the \textit{Dynamical Matrix} ($\mat{D}$):
\begin{equation}
    \label{eq:dyn}
    U - U_0 \approx \frac{1}{2} \vec{u}^T \mat{D}^0\vec{u} = \sum_{ij}^{N} D^0_{ij} u_i u_j ,
\end{equation}
where
\begin{equation}
    D^0_{ij} = \frac{\partial^2 U}{\partial \vec{r}_i \partial \vec{r}_j} \Bigg|_{\vec{\delta}_0}
\end{equation}
is the Dynamical Matrix (or Hessian) at equilibrium and depends on the intrinsic material properties and arrangements of the particles.
The elements of $\mat{D}$ for the one-dimensional chain can be obtained using  the pairwise potentials (\cref{eq:potential}) as follows: 
\begin{equation}
\begin{aligned}
    i \neq j: \quad & D_{ij} = - \frac{\epsilon(\alpha-1)}{\sigma_{ij}^2} {\left(1-\frac{r_{ij}}{\sigma_{ij}}\right)}^{\alpha-2} \left(\frac{x}{r_{ij}}\right)  \\
    & \qquad + \frac{\epsilon}{r_{ij} \; \sigma_{ij}} {\left(1-\frac{r_{ij}}{\sigma_{ij}}\right)}^{\alpha-1} \left(1-\frac{x}{r_{ij}}\right); \\
    i = j: \quad & D_{ij} = - \sum_{i \neq j} D_{ij}.
\end{aligned}
\end{equation}
For a system with $d$ dimensions and $N$ particles, $\mat{D}$ will be a $dN$-dimensional matrix. 
So, for the granular chains in this paper, $\mat{D}$ is an $N \times N$ symmetric matrix. 

If the harmonic approximation holds (for small amplitude dynamic displacements), the equations of motion of the unforced ($\vec{F}_{ext}=\vec{0}$) Hamiltonian system are obtained by differentiating \cref{eq:dyn} with respect to time:
\begin{equation}
\frac{\partial U}{\partial \vec{r}}  \frac{d \vec{u}}{dt} = \frac{1}{2}  \frac{d\vec{u}^T}{dt}  \mat{D}^0  \vec{u} + \frac{1}{2}  \vec{u}^T  \mat{D}^0 \frac{d\vec{u}}{dt},
\end{equation}
and noting that $\mat{D}^0$ is symmetric and 
\begin{equation}
    -\frac{\partial U}{\partial \vec{r}} = \mat{M} \frac{d^2 \vec{u}}{dt^2},
\end{equation}
where $\mat{M}$ is the mass matrix in which the diagonal elements are the mass of the corresponding particle ($M_{ii} = m$),
we arrive at the equations of motion,
\begin{equation}
\mat{M}\frac{d^2 \vec{u}}{dt^2} +\mat{D}^0 \vec{u} = 0.
\label{eq:eom}
\end{equation}

\cref{eq:eom} is a system of $N$ homogeneous second-order differential equations and can be solved by finding the eigenvalues ($\omega_k$) and eigenvectors ($\vec{e}_k=[e_1^k, e_2^k, ..., e_N^k]$) of the Dynamical Matrix. The eigenvalues are the Linear Normal Modes of the system, and provide the periodic vibrational response of the form $u_i(t) = \sum_k^N \alpha^k_i \hat{e}_i^k e^{i\omega_kt+\phi_k}$, where $\hat{e}_k$ is the normalized eigenvector (shape of the mode), and $\vec{\alpha}_k=[\alpha_1^k, \alpha_2^k, ..., \alpha_N^k]$ (amplitude of the mode) and $\phi_k$ are determined by the $2N$ initial positions and velocities when the system is disturbed from the equilibrium position. 
Assuming  the initial velocities are zero ($v_i(0)=0$), we have $\phi_k=0$ and $\vec{u}(0) = \sum_k^N \alpha_k \hat{e}_k$, and the positions and velocities are
\begin{align}
    \label{eq:harmonic}
    u_i(t) = & \sum_k^N {\alpha_i^k \cos{(\omega_k t + \phi_k)} \hat{e}^k_i}, \\
    v_i(t) = & - \sum_k^N {\alpha_i^k \omega_k \sin{(\omega_k t + \phi_k)} \hat{e}^k_i}.
\end{align}

The approach outlined here is similar to the spectral analysis of lattice dynamics with the harmonic balance method to obtain a numerical estimate of the periodic solutions to the nonlinear equations of motion \cite{Maradudin1971}.

\subsection{Numerical Simulation}
\label{sec:numericalSim}
In this paper, we use the Discrete Element Method (DEM) \cite{Cundall.Strack1979} to simulate the motion of the interacting particles in a granular crystal. The main steps of DEM are presented in \cref{alg:DEM} in \cref{sec:appendix} \footnote{Modified from \cite{FatihGoncu2012}.}; the simulation starts with the initial configuration and updates the positions and velocities by numerically integrating the equations of motion (\cref{eq:motion}). 

Since our granular packings are made of particles with various material properties and are initially compressed with a uniform force, we need to ensure that the initial configuration is statically stable (the ground state $\vec{u}=\vec{0}$ and $\dot{\vec{u}}=\vec{0}$ is the minimum of energy). Here, we adopt a packing generation protocol that applies successive compression/decompression by changing the particle sizes \cite{Gao.etal2009a, Franklin.Shattuck2016}. An energy minimization technique, Fast Inertial Relaxation Engine (FIRE) \cite{EcheverriRestrepo.etal2013}, is used to relax the interparticle forces and reach an equilibrium state by repetitive particle deformations. With this method, we can find the particles' initial positions for a mechanically stable configuration with a given boundary condition \cite{Asenjo-Andrews2013}.

To integrate the equations of motion, we use a Velocity-Verlet time integration scheme (\cite{Verlet1967}) similar to the time-synchronized Leapfrog method. This method which is based on computing the half-step velocity ($\vec{v}(t+\frac{\Delta t}{2}) = \dot{\vec{r}}(t) + \frac{1}{2} \ddot{\vec{r}}(t) \Delta t$) is presented in \cref{alg:verlet} in \cref{sec:appendix}.

\section{Koopman Theory and Data-driven Koopman}
\label{sec:koopmanTheory}
Aside from the limitations mentioned in \cref{sec:relatedWorks}, the classic dynamical analysis methods are also highly dependent on the accuracy of the underlying mathematical model, limiting their applications in many real-world systems. Aspects of the physical system that are not captured by the derived dynamical equations, such as dissipation effects and discrepancies in the fabrication of components, can lead to significant deviation from the predicted behavior. 
Therefore, despite all the studies mentioned at the start of this paper, we still lack a fundamental understanding of the role of distinct sources of nonlinearity and disorder in the anharmonic response of granular media \cite{Porter.etal2015}.

Among the data-driven methods for the analysis and modeling of underlying nonlinear systems, operator-theoretic frameworks provide interpretable and yet accurate descriptions of the global dynamics of nonlinear systems \cite{Nandanoori.etal2022}. Unlike the geometric approaches (in state space) discussed in the previous section, operator-theoretic analysis is done in the function space and can (in the case of the Koopman operator) globally linearize the nonlinear dynamics, thus making it possible to apply linear analysis and control methods to the nonlinear system.

\subsection{Koopman Theory}
\label{sec:koopmanDetails}
The key idea in the Koopman operator theory {Here we discuss the modern framework. The original Koopman theory was introduced in $1931$ \cite{Koopman1931}. It was then extended to systems with continuous eigenvalue spectrum \cite{Koopman.Neumann1932}, and dissipative systems \cite{Mezic.Banaszuk2004, Mezic2005}.} is the usage of special basis functions to lift the dynamics of a finite-dimensional nonlinear system and map it to an infinite-dimensional \textit{Hilbert} space where the original system becomes linear \cite{Brunton.etal2022}. Consequently, since linear systems are completely characterized by their spectral decomposition, a modal decomposition in this linear functional space can extract the global characteristics of the nonlinear system without requiring local linearizations.

In this paper\footnote{The mathematical notation in this section is adapted from \cite{Schmid2022, NathanKutz.etal2018, Takeishi.etal2017, Brunton.etal2022, Lusch.etal2018, Kaiser.etal2021}.}, we consider a discrete-time autonomous system $x_{t+1}=F(x_t)$, where $x$ is the system's state variable and $F$ is the dynamics that governs the evolution of the state in time. While $F$ is a nonlinear function, the Koopman operator $\mat{K}$ is defined as a linear operator that acts on the observable functions\footnote{Or measurement functions.} ($y=g(x)$) of the state such that $\mat{K} g(x_t) = g \circ F(x_t) = g(x_{t+1})$\footnote{While $F$ governs state $x$ in time, Koopman operator governs the evolution of function $g$ in the infinite-dimensional Hilbert space.}, where $\circ$ denotes the composition operation. Since $\mat{K}$ is assumed to be linear, it can be decomposed to a set of eigenfunctions $\{\varphi_1, \varphi_2, ...\}$ and eigenvalues $\{\lambda_1, \lambda_2, ...\}$ such that $\mat{K} \varphi_i = \lambda_i \varphi_i$. The superposition principle is valid in this linear functional space, so the observable $g$ can be written as a linear combination of the eigenfunctions, i.e. $g(x) = \sum_{i=1}^{\infty} c_i \varphi_i(x)$ where $\{c_1, c_2, ...\}$ is an infinite set of coefficients.

If we start from $g(x_0)$ and apply the Koopman operator to each side of the equation we will obtain
\begin{equation}
    \mat{K} g(x_0) = g(x_1) = \mat{K} \sum_{i=1}^{\infty} c_i \varphi_i(x_0)=\sum_{i=1}^{\infty} c_i \lambda_i \varphi_i(x_0).
\end{equation}
Repeating this procedure will result in the modal decomposition of the observable function at time $t$:
\begin{equation}
    \label{eq:koopman}
    g(x_t) = \mat{K}^t g(x_0) = \sum_{i=1}^{\infty} c_i \lambda_i^t \varphi_i(x_0),
\end{equation}
where $c_i$ is known as the $i$th Koopman mode (or dynamic mode) corresponding to $i$th Koopman eigenfunction $\varphi_i$ (or intrinsic coordinates) and $i$th Koopman eigenvalue $\lambda_i$\footnote{$\lambda_i=\mu_i + j \omega_i$ is a complex number, $\angle \lambda_i$ and $|\lambda_i|$ determine the frequency and decay/growth rate of a given mode.}. The sequence of ${\{(\lambda_i, \varphi_i, c_i)\}}_{i=1}^{\infty}$ is called the Koopman mode decomposition \cite{Mezic2005}. In this context, the Koopman operator transforms the system's coordinates in space and time to linear space, enabling the characterization of the observables through the spectral properties of the dominant Koopman modes\footnote{Koopman eigenfunctions are intrinsic coordinates, along which the dynamics are linear.}. 

According to the Koopman theory, $\mat{K}$ is infinite-dimensional, but one can obtain a finite-dimensional representation by restricting $\mat{K}$ to a Koopman-invariant subspace spanned by a finite set of (nonlinear) observable functions \cite{Brunet.etal2008}. Dynamic Mode Decomposition (DMD) is the leading data-driven method that approximates the Koopman operator by assuming $g(x)=x$ \cite{NathanKutz.etal2018}. 
Using DMD for analyzing complex systems from direct measurements has attracted a lot of interest over the past decade and many improvements and extensions have been proposed such as extended DMD (EDMD) \cite{Williams.etal2015}, DMD with control (DMDc) \cite{Proctor.etal2016}, and physics-informed DMD (piDMD) \cite{Baddoo.etal2023}.

Representing the nonlinear dynamics in a linear framework provides the opportunity to apply numerous tools and concepts developed for linear systems analysis. Koopman eigenfunctions represent the global phase portrait of the nonlinear system and Koopman modes can be used to identify different dynamical regimes in a system. In the case of direct measurement of the system ($g(x)=x$), the Koopman modes can also have a physical interpretation \cite{Nandanoori.etal2022}. For example, with spatial measurements of the system, Koopman modes are spatial modes that have the same temporal dynamics, meaning that the system components oscillate together. Aside from the characterization of complex dynamics, Koopman operator theory can be used for making predictions about future system states. Moreover, the linear Koopman embedding of a nonlinear system makes it possible to apply linear control methods to nonlinear dynamical systems.

\subsection{Data-driven framework}
\label{sec:dataDriven}
Defining an adequate set of observable functions ($\{g(x)\}$) that can form a Koopman invariant subspace is a challenging task, requiring expert knowledge of the underlying system. Simple function dictionaries (such as identity, polynomials, etc.) may necessitate higher dimensional Koopman subspace, and more complex functions (e.g. using kernel methods) may make the Koopman embedding less interpretable. An alternate method is to find the Koopman eigenfunctions directly from the state measurements. An emerging research direction is the usage of Deep Neural Network architectures to represent the eigenfunctions \cite{Lusch.etal2018}. With this direct approach, we'll replace $g(x_t)=x_t$ in the equations from the previous section, and so the Koopman eigenfunctions ($\varphi_i$) will satisfy
\begin{equation}
    \label{eq:deepkoopman}
    \varphi(x_{t+1}) = \mat{K} \varphi(x_t) = \lambda \varphi(x_t).
\end{equation}

\begin{figure*}[htp!]
\centering
\includegraphics[width=1.0\textwidth]{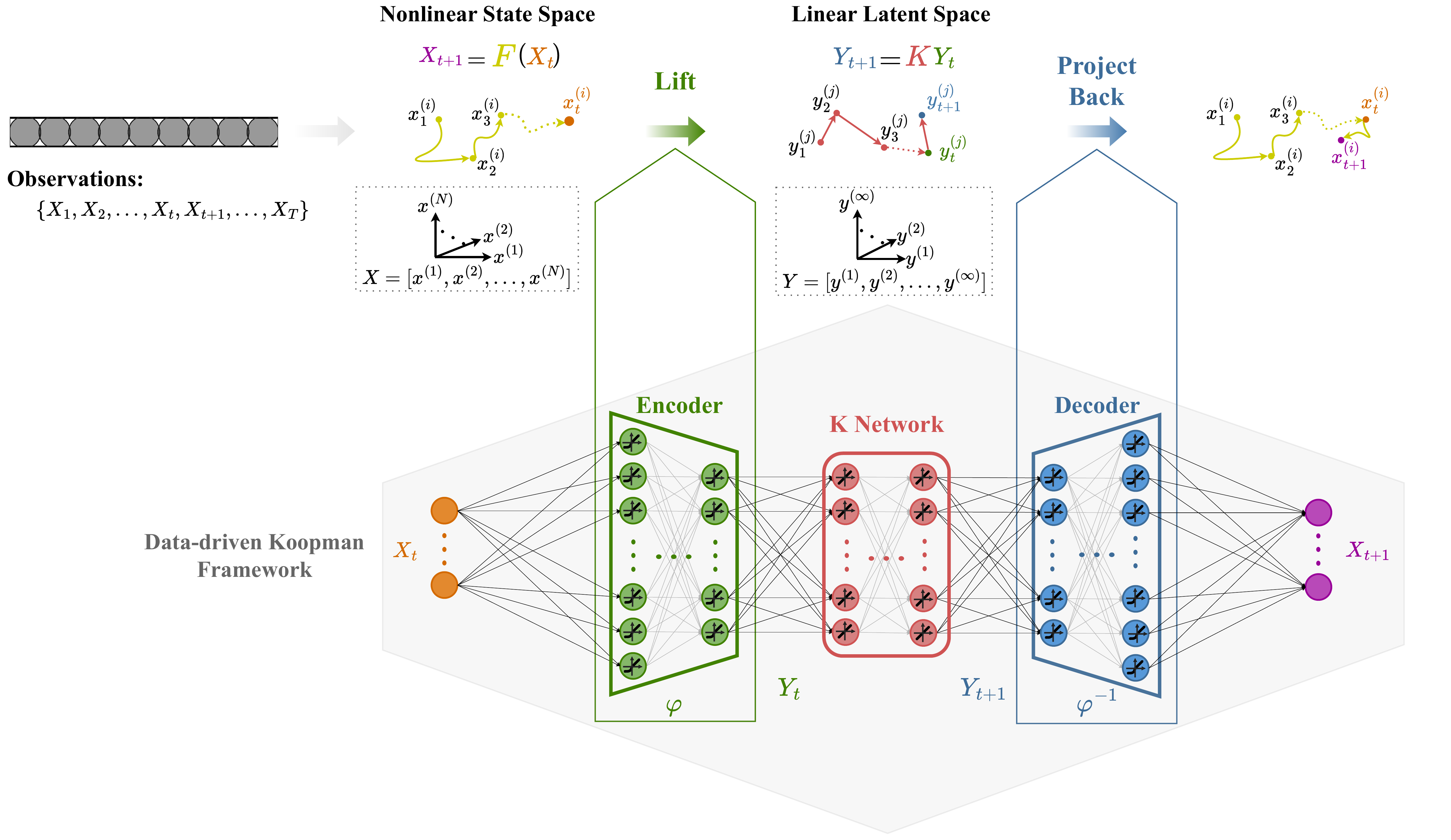}
\caption{\label{fig:koopman}Deep neural network architecture for approximating the Koopman operator. The encoder network learns the Koopman embedding ($\Vec{Y}_t$) from the measurements of the system's states ($\Vec{X}_t$). The \textit{K Network} predicts the latent state in the next time step ($\Vec{Y}_{t+1}$). The loss function (\cref{eq:koopmanloss}) is designed to ensure that the physical constraints of the dynamical system (prediction loss over the input trajectories) and mathematical assumptions of the Koopman theory (reconstruction loss in the encoder/decoder networks and linearity of \textit{K network}) are met.
}
\end{figure*}

\cref{fig:koopman} shows the common Deep Neural Network architecture for learning the Koopman eigenfunctions\footnote{$\varphi$ network will represent the Koopman eigenfunctions if we assume $\mat{K}$ is diagonalized.} from measurements of a system. Here, the system is a granular crystal and the system state is the displacements of the particles during the simulation time ($X_t=[r_0, r_1, ..., r_N]$). The encoder network maps the system state to intrinsic coordinates ($y_t=\varphi(x_t)$), the $\mat{K}$ network (Koopman operator) advances the coordinates to the next time step ($y_{t+1}=\mat{K} y_t$), and the decoder network maps $y_{t+1}$ back to the system state $x_{t+1}$. To enforce the linearity of the Koopman operator, the $\mat{K}$ network has linear layers with no bias terms. The dimension of the embeddings ($y$) and the number of layers in each network are hyperparameters that have to be determined by the knowledge of the system.

In this paper, we utilize the method presented in \cite{Lusch.etal2018} where three loss functions are defined to enforce the physical constraints of the system and the assumptions made by the Koopman theory:
\begin{enumerate}
    \item Reconstruction: 
    
    $\mathcal{L}_{recon} = \Big\| x_t-\varphi^{-1}(\varphi(x_t)) \Big\|_{MSE}$
    
    \item Linearity (over $T$ time steps):
    
    $\mathcal{L}_{lin} = \frac{1}{T} \sum_{t=0}^{T} \Big\| \varphi(x_{t})-\mat{K}^t \varphi(x_0) \Big\|_{MSE}$
    
    \item Prediction (over $P$ time steps):
    
    $\mathcal{L}_{pred} = \frac{1}{P} \sum_{t=0}^{P} \Big\| x_{t}-\varphi^{-1}(\mat{K}^t \varphi(x_0)) \Big\|_{MSE}$
\end{enumerate}

Here $T$ is the length of the training trajectories, and $P$ is a hyperparameter indicating the number of time steps for the state prediction. So the total loss is
\begin{equation}
    \label{eq:koopmanloss}
    \mathcal{L} = \alpha_1 \mathcal{L}_{recon} + \alpha_2 \mathcal{L}_{pred} + \alpha_3 \mathcal{L}_{lin}.
\end{equation}

For the results presented in the next section, we used an open-source package called \textit{NeuroMANCER} \cite{Drgona.etal2023} that provides PyTorch implementations of the Deep Koopman operator with and without control. The details of the architecture and simulation parameters are presented in \cref{sec:modelingUnforced}.

\section{Modeling the Unforced Granular Chain}
\label{sec:modelingUnforced}
\subsection{Implementation Details}
\label{sec:unforcedDetails}
In this section, we consider a two-particle granular chain with the simulation parameters presented in \cref{tab:KP_sim}. The goal is to apply the deep Koopman framework described in \cref{sec:koopmanTheory} to this system and analyze the intrinsic nonlinear dynamics of the system. \cref{tab:KP_params} includes the parameter values of the encoder/decoder and Koopman network. We generate a dataset with the particles' trajectories in an unforced system to train the networks. The system is simulated with the Discrete Element Method described in \cref{sec:numericalSim}.
\begingroup
\setlength{\tabcolsep}{8pt}
\begin{table}[h]
\caption{Deep Koopman Network Parameters.}
\label{tab:KP_params}
\begin{tabular}{l|cc}
\toprule
    \bfseries Parameter & \bfseries Encoder/Decoder & \bfseries K Network \\
\midrule \midrule
    Input Size & $ny=4$ & $50$\\
    Hidden Layer & $6 \times [200]$ & \textit{None}\\
    Output Size & $50$ & $50$ \\
    Nonlinearity &  \textit{ELU} &  \textit{None}\\ 
    Bias & \textit{True} & \textit{False} \\
\bottomrule
\end{tabular}
\end{table}
\endgroup
To capture the full dynamics of the system, we need to have a sufficiently large dataset with trajectories that cover the full phase space. Therefore, we randomly set the particles' initial displacements within a given range (see \cref{tab:KP_sim}) and generate $10^3$ trajectories by recording the particle displacements and velocities in time. It is worth mentioning that in this experiment, we assume the particles have zero initial velocity. In the DEM simulations, we need a small time step to ensure the numerical integration is converging to its true value. Therefore, the number of data points in each trajectory is large and the training will be time-consuming. We resample the trajectories with a predefined sampling rate to reduce the training load. The dataset is then split into training, validation, and test sets, and the network is trained with the setup presented in \cref{tab:KP_train}.
\begingroup
\setlength{\tabcolsep}{8pt}
\begin{table}[h]
\caption{Simulation Setup for Koopman Analysis.}
\label{tab:KP_sim}
\begin{center}
\begin{tabular}{l|c}
\toprule
    \bfseries Parameter & \bfseries Value \\
\midrule \midrule
    Simulation Time ($T$) & $1e3$ \\
    Time Step ($\Delta t$) & $5e-3$ \\
    Sampling Rate & $500$ \\
    Damping ($B, B_{pp}, B_{pw}$) & $0$ \\
    Stiffness Ratio & $2.3$ \\
    Particle Mass ($m$) & $1$ \\
    Particle Diameter ($\sigma_0$) & $1e-1$ \\
    Particle Contacts & \textit{Hertzian} \\
    Packing Fraction ($\phi$) & $1e-1$ \\
    Relaxation Time ($FIRE$) & $1e6$ \\
    Initial Displacement ($x_0$) & $\in [-1e-2, 1e-2]$ \\
    Initial Velocity ($v_0$) & $0$ \\
    Number of Trajectories & $10e3$ \\
\bottomrule
\end{tabular}
\end{center}
\end{table}

\begin{table}[h]
\caption{Training Setup for Koopman Analysis.}
\vskip -0.1in
\label{tab:KP_train}
\begin{center}
\begin{tabular}{l|c}
\toprule
    \bfseries Parameter & \bfseries Value \\
\midrule \midrule
    Training Set & $500$ samples \\
    Validation Set & $500$ samples \\
    Test Set & $100$ samples \\
    Prediction Window ($P$) & $500$ steps \\
    Optimizer & \textit{Adam} \\
    Learning Rate & \textit{1e-4} \\
    Epochs & $2e4$ \\
\bottomrule
\end{tabular}
\end{center}
\vskip -0.2in
\end{table}
\endgroup

\subsection{Results}
\label{sec:unforcedResults}
\begin{figure}[htp!]
\centering
\includegraphics[width=1.0\columnwidth]{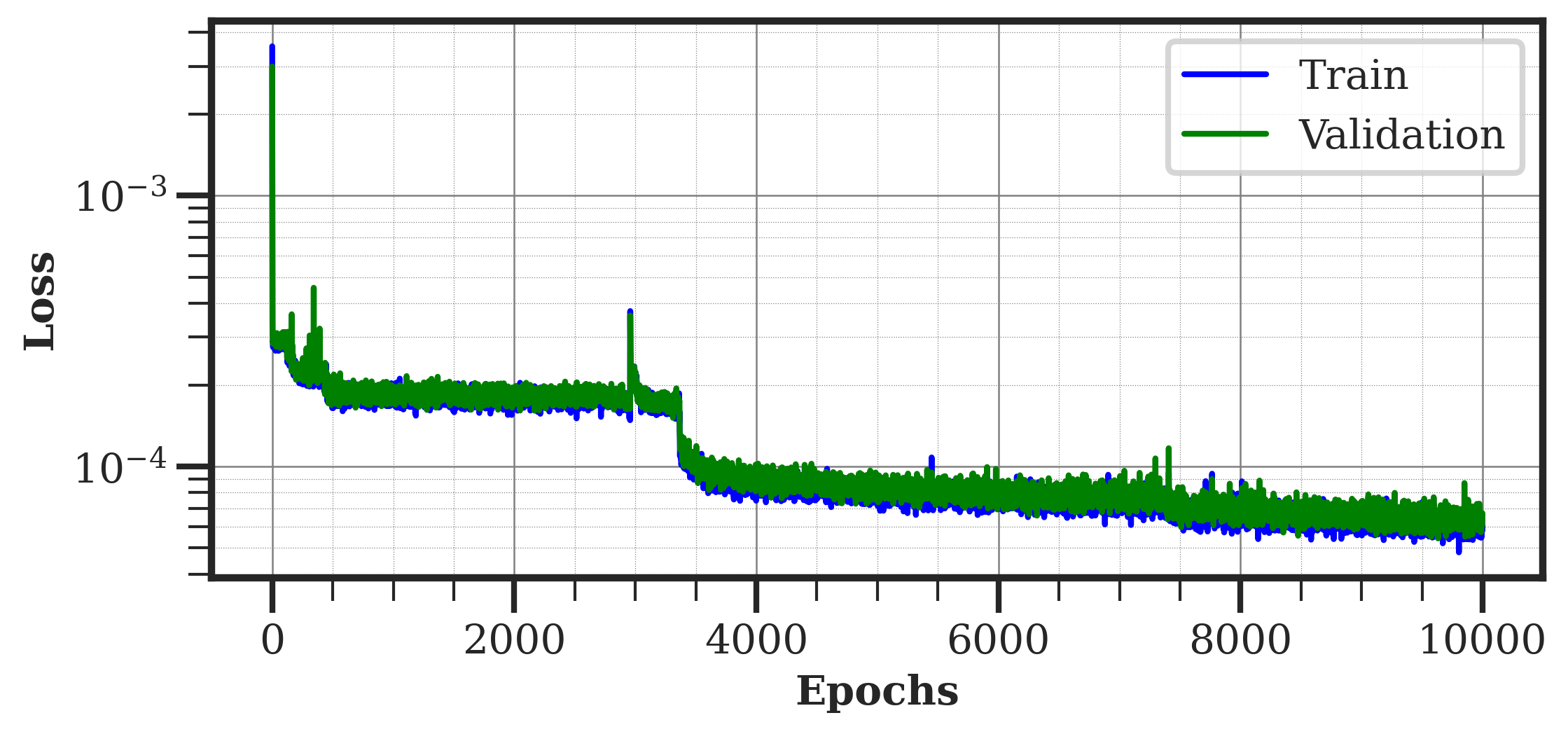}
\caption{\label{fig:DK_loss}Value of the loss function \cref{eq:koopmanloss} over the training and validation datasets. The deep Koopman network is trained on the displacement and velocity trajectories obtained from a numerically simulated two-particle system. The particles' initial positions are selected randomly from a predefined range indicated in \cref{tab:KP_sim}.
}
\end{figure}

As noted in \cref{sec:dataDriven}, the loss function has three terms including reconstruction $\mathcal{L}_{recon}$, prediction $\mathcal{L}_{pred}$ and linearity $\mathcal{L}_{lin}$ which are defined in \cref{eq:koopmanloss}. We set the coefficients as $\alpha_1, \alpha_2, \alpha_3=1$. \cref{fig:DK_loss} shows the training and validation loss during the training.
\begin{figure}[htp!]
\centering
\includegraphics[width=1.0\columnwidth]{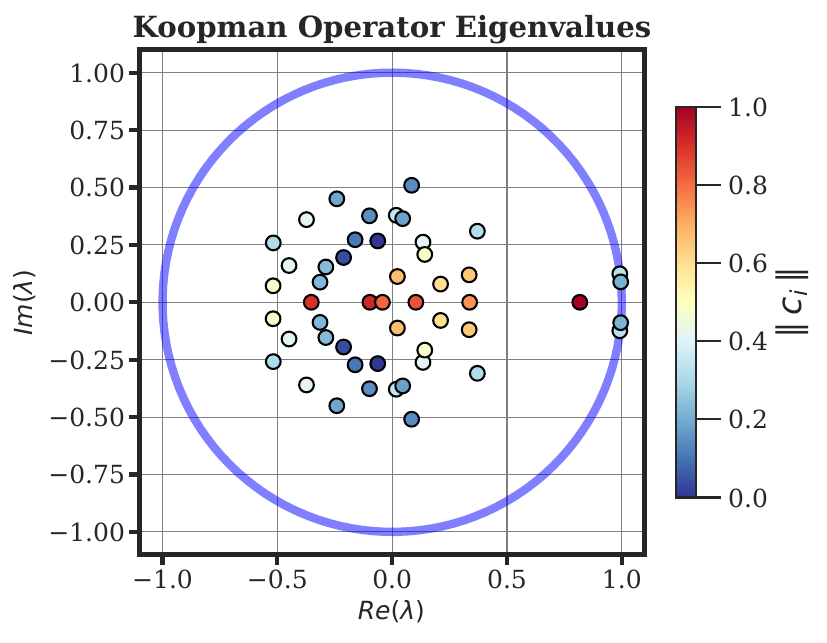}
\caption{\label{fig:DK_dominance}Koopman eigenvalues for the unforced two-particle system. The eigenvalues determine the frequency of the Koopman modes ($Im(\lambda)$) and their decay/growth rate ($Re(\lambda)$). The blue outline shows the unit circle, and the color of each mode indicates its dominance ($\parallel c_i \parallel$). 
}
\end{figure}

\begin{figure*}[htb!]
\centering
\includegraphics[width=1.0\textwidth]{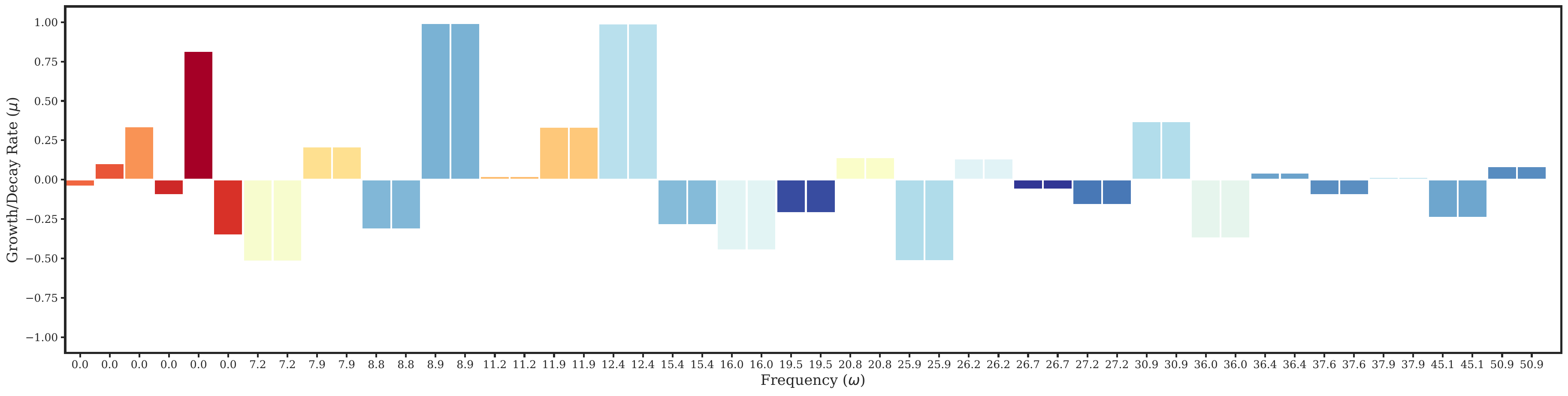}
\caption{\label{fig:DK_rate}Growth/decay rate of the Koopman modes for the unforced two-particle system. The color indicates the dominance of the modes ($\parallel c_i \parallel$) corresponding to \cref{fig:DK_dominance}. The horizontal axis shows the frequency of the modes ($\omega=\log(\lambda)/\Delta t$). The first four modes are background modes and have a frequency of zero. Moreover, for each nonzero frequency, there are two complex conjugate modes with the same oscillation rate.
}
\end{figure*}
\begin{figure*}[htp!]
\centering
\includegraphics[width=1.0\textwidth]{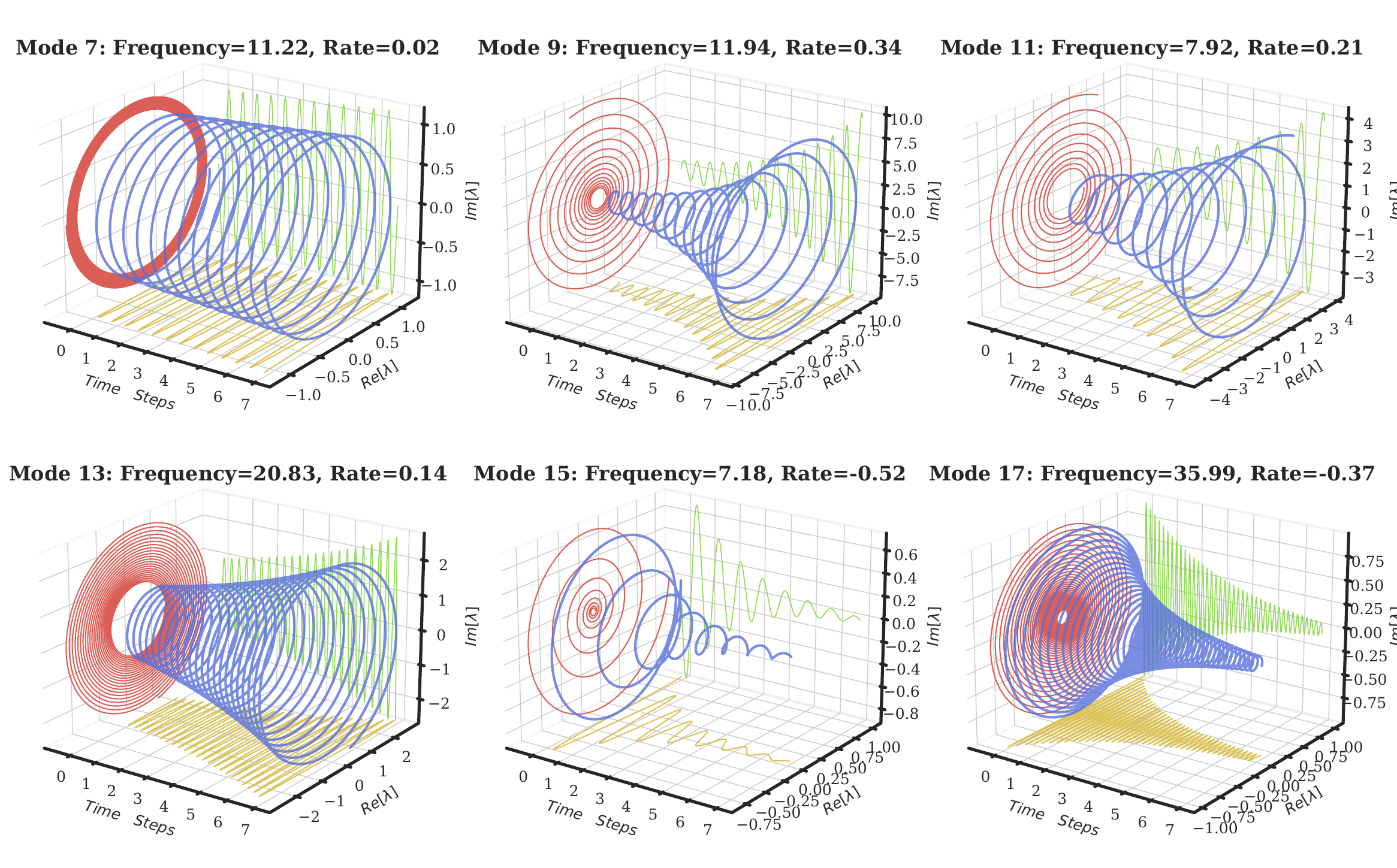}
\caption{\label{fig:DK_modes}The top six nonzero dominant Koopman modes for the two-particle system. In each panel, the blue graph shows the mode in the complex plane, where the real and imaginary parts of the mode are projected onto each face of the 3D plot. 
}
\end{figure*}

After training the network, we can derive the Koopman mode decomposition including eigenvalues, eigenfunctions, and dynamic modes. The Koopman eigenvalues ($\lambda$) are shown in the complex plane in \cref{fig:DK_dominance} with the colors indicating the value of $c_i$, which quantifies the dominance of each mode in the observed data. \cref{fig:DK_rate} shows each mode's decay/growth rate with the frequencies indicated on the horizontal axis.

Since symmetry in the dynamics can give rise to background modes with an eigenvalue of zero, as we see in \cref{fig:DK_dominance} there are four zero modes in our system. The top six nonzero dominant modes (excluding the complex conjugates) are shown in \cref{fig:DK_modes}.

\begin{figure*}[htp!]
\centering
\includegraphics[width=1.0\linewidth]{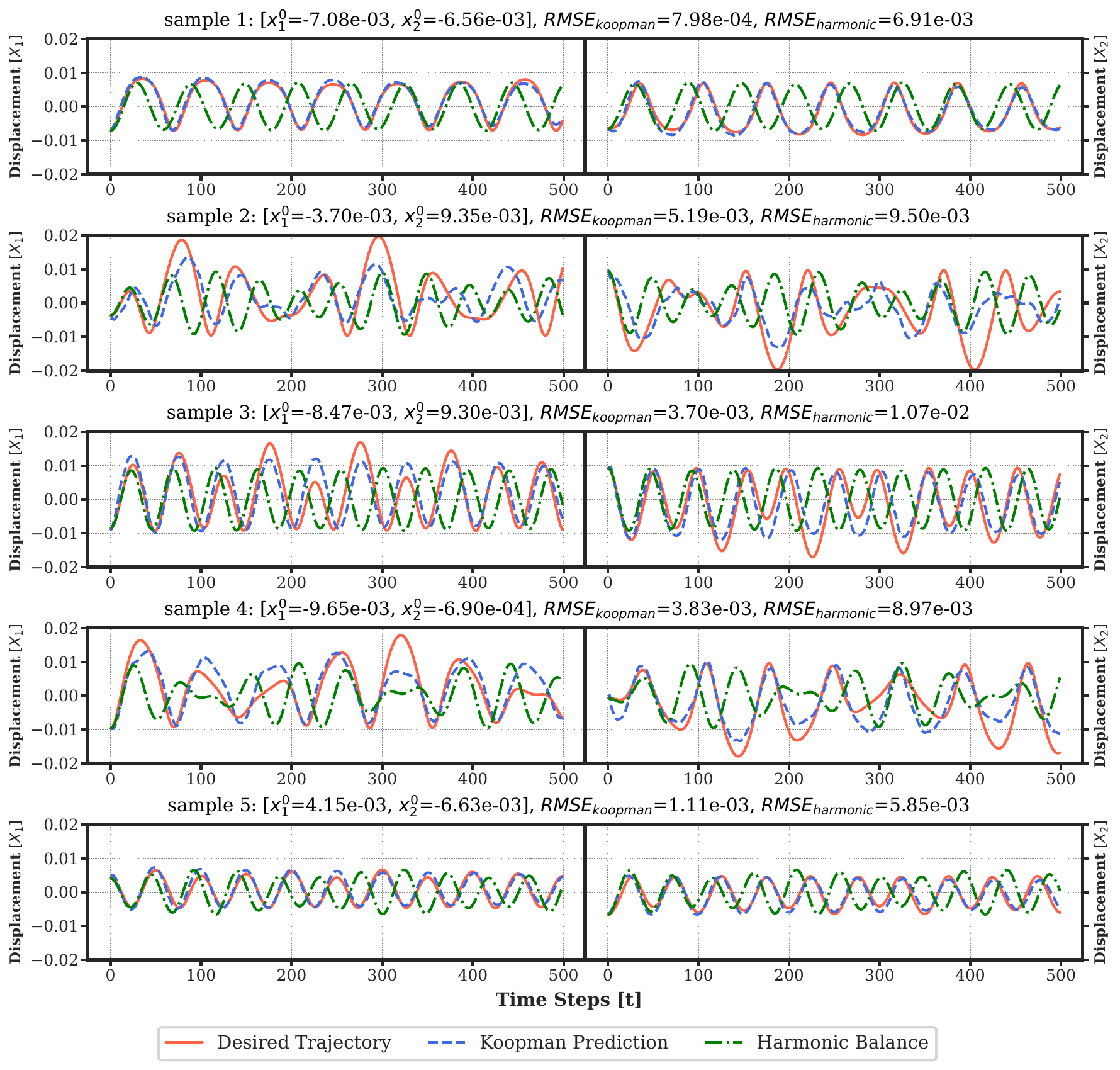}
\caption{\label{fig:DK_predictions}
Desired and predicted trajectories of the particles for five different initial conditions. Each panel shows one sample from the test dataset with the initial conditions (particle displacements $X^0=[x^0_1, x^0_2]$) indicated above it. The plots in each panel show the displacement of the first (left) and second (right) particles in time. Solid lines (red) show the true trajectories obtained from the numerical simulations. The dashed lines (blue) show the predictions from the Koopman framework. The dot-dashed lines (green) show the results of harmonic approximation. The RMSE (\cref{eq:RMSE}) is printed on top of each panel.
}
\vskip -0.1in
\end{figure*}
To evaluate the network performance, we selected five sample trajectories from the test set and assigned the initial particle positions and velocities to the encoder network. Using the trained Koopman network, we can predict the particle positions and velocities in the subsequent time steps and compare them to the reference trajectory obtained from the numerical simulation. \cref{fig:DK_predictions} presents reference and predicted trajectories for these five test samples. 

\begin{figure*}[htp!]
\centering
\includegraphics[width=1.0\linewidth]{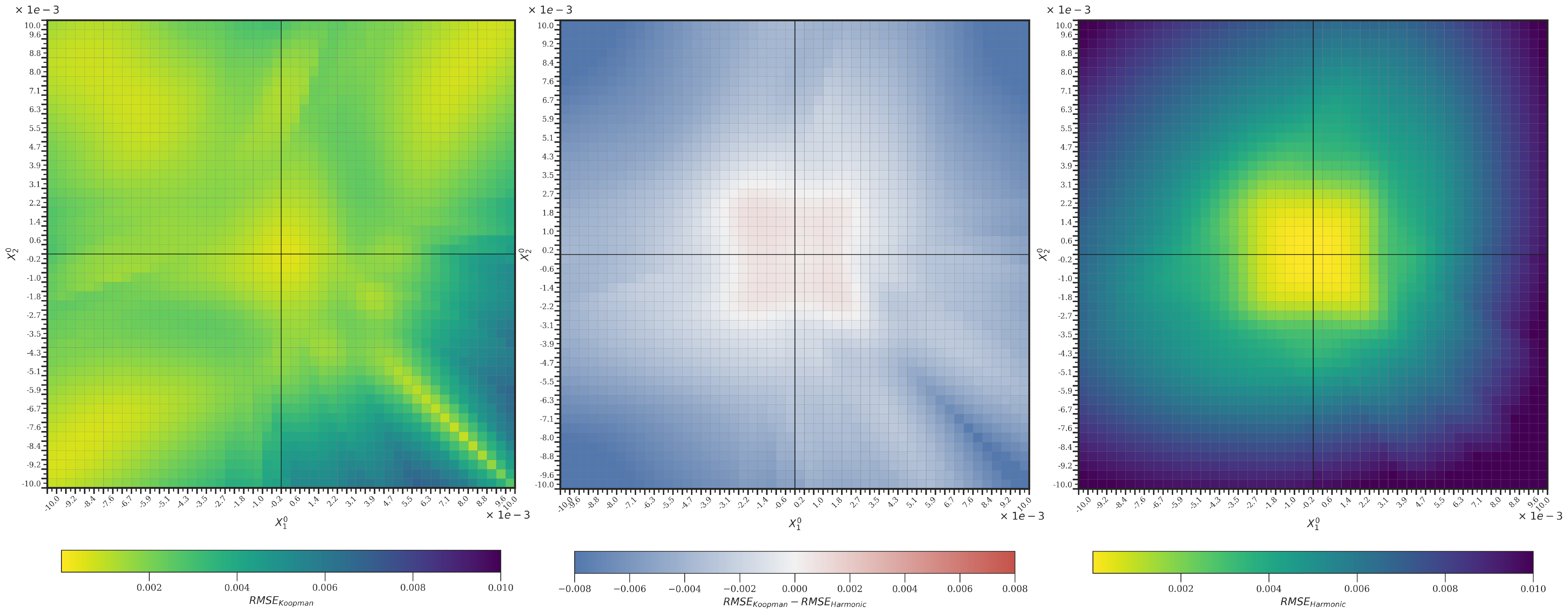}
\caption{\label{fig:DK_predictions2}Accuracy of Koopman network predictions versus harmonic approximation. The horizontal and vertical axes in each panel represent the displacement of the first ($X^0_1$) and second ($X^0_2$) particles in a two-particle system. Root Mean Squared Error (\cref{eq:RMSE}) between the desired and predicted trajectory is calculated for equally placed data points on the horizontal and vertical axes. The panel on the left shows that the Koopman framework provides high prediction accuracy for a wider range of particle displacements, whereas harmonic approximation (the right panel) can only predict the particles' displacement in the (weakly) nonlinear regime when the initial displacement of the particles is small. The panel in the middle shows the difference between the two prediction errors.
}
\end{figure*}

To compare the accuracy of the Koopman network's predictions and harmonic approximation, we also calculated the \textit{Root Mean Squared Error (RMSE)} of each case using the following formula:
\begin{equation}
    \label{eq:RMSE}
    \mathit{RMSE}(X, \hat{X}) = \sqrt{\frac{\sum_{i=1}^{N}{(X_i-\hat{X_i})}^2}{N}}, 
\end{equation}
where $X$ indicates the real trajectory, $\hat{X}$ is the prediction, and $N$ indicates the number of particles in the granular chain, which is considered two in our experiments. As we can see in \cref{fig:DK_predictions2}, for small particle displacements, the system is weakly nonlinear, and harmonic approximation predicts particles' displacements with high accuracy.

\section{Discussion}
\label{sec:discussion}
\begin{figure*}[htp!]
\centering
\includegraphics[width=1.0\linewidth]{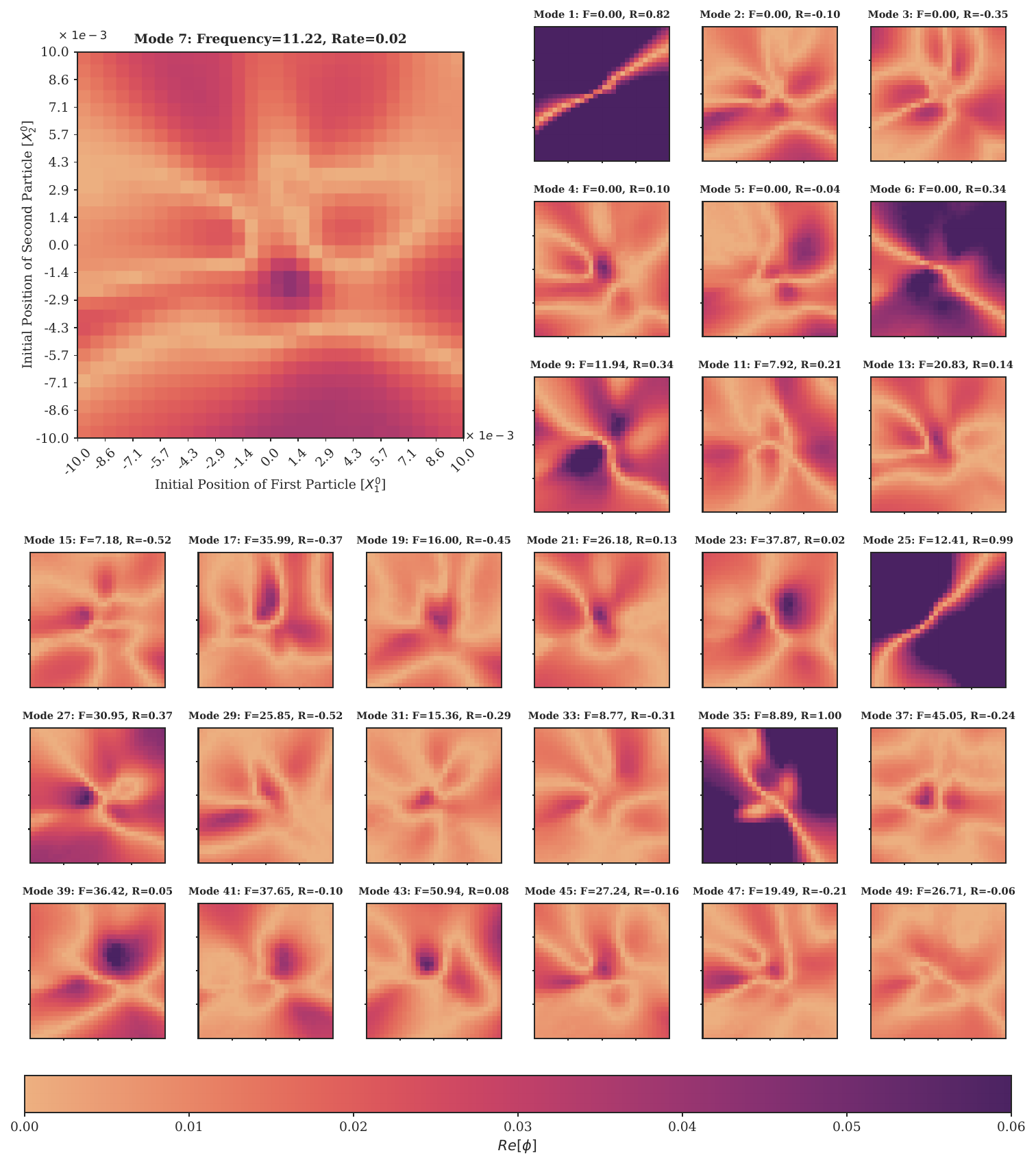}
\caption{\label{fig:DK_eigenfuncs_real}
The real component of the Koopman eigenfunctions for the two-particle system. Each panel corresponds to one of the Koopman modes extracted from the trained network. The axes in each panel represent the initial conditions (particles' initial displacements) $X_0^1$ and $X_0^2$. The color indicates the contribution of each mode to the system's dynamics at the selected initial condition.
}
\end{figure*}

\begin{figure*}[htp!]
\centering
\includegraphics[width=1.0\linewidth]{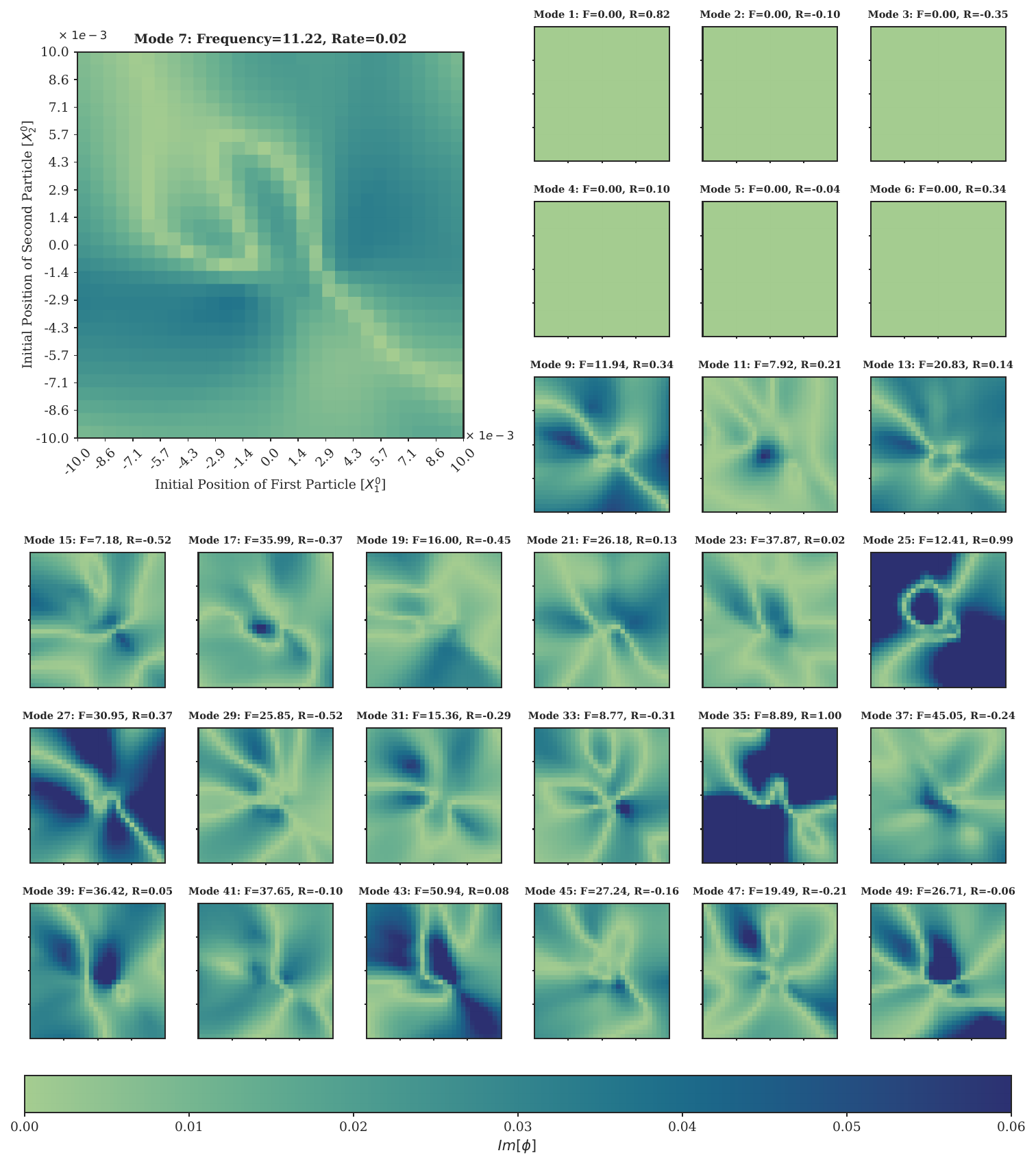}
\caption{\label{fig:DK_eigenfuncs_img}
The imaginary component of the Koopman eigenfunctions for the two-particle system. Each panel corresponds to one of the Koopman modes extracted from the trained network. The axes in each panel represent the initial conditions (particles' initial displacements) $X_0^1$ and $X_0^2$. The color indicates the contribution of each mode to the system's dynamics at the selected initial condition.
}
\vskip 0.2in
\end{figure*}

After obtaining the Koopman mode decomposition, we can analyze the system's dynamical regimes and their stability by examining the real and imaginary components of the eigenfunctions. To visualize these global dynamics, we produced $900$ initial conditions from equally-placed points in $[-1e-2, 1e-2]$ as the initial displacements of the first and second particles. The real and imaginary components of the eigenfunction evaluated at each initial condition ($\varphi_i(X^0)$) provide an overview of the phase space which can later be used to identify the dynamical regimes in the system. \cref{fig:DK_eigenfuncs_real} and \cref{fig:DK_eigenfuncs_img} present these components.

Inspecting these plots can provide information about the stability of each trajectory (starting from each initial condition) and the existing harmonic components. Further, we can identify various dynamical regimes, fixed points, and bifurcations in the system. 

\begin{figure*}[htp!]
\centering
\includegraphics[width=1.0\linewidth]{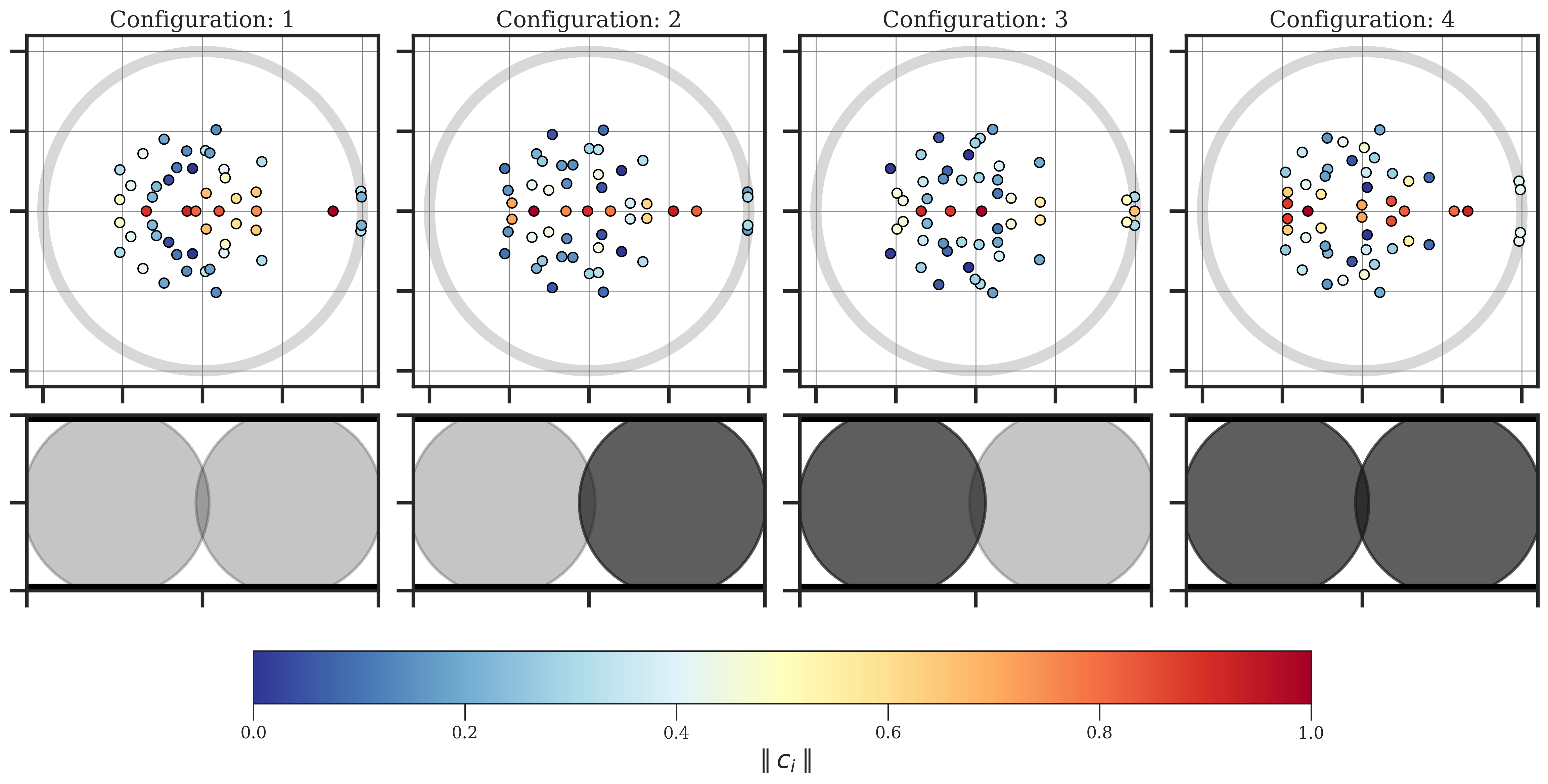}
\caption{\label{fig:DK_compare}
Koopman modes in four distinct particle configurations. The light/dark grey indicates the soft/stiff particles in each of the four panels. For each configuration, Koopman modes are plotted in the complex plane, where the color of each point quantifies the dominance of the mode($\parallel c_i \parallel$).
}
\end{figure*}

The deep Koopman framework presented here maps the system dynamics to a linear space characterized by the Koopman modes. This enables us to compare different systems at the level of their dynamics by probing their Koopman modes. Here, we considered four two-particle systems with distinct configurations of soft and stiff particles (see \cref{fig:DK_compare}). Investigating the Koopman modes in each case might provide information about how structural properties affect the system dynamics in terms of the frequencies of the dominant modes present in the linear latent space. 

\section{Conclusions}
\label{sec:conclusions}
This paper started with an overview of the static and dynamic characteristics of granular crystals, highlighting the nontrivial nonlinear wave responses that can emerge in such systems. Then, we provided an overview of a data-driven method based on the Koopman theory that can capture the global dynamics of the system without resorting to any simplifying assumptions on the physics model or any approximations of the nonlinear dynamics. Such a data-driven framework can provide several advantages, including:
\begin{itemize}
    \item \textit{Capturing the system's dynamics without relying on the physics model:} since the network is trained on system observations, one can incorporate the measurements from the hardware setup instead of numerical simulations and thus extract the dynamics without any particular assumptions on the system parameters or the physics model.
    
    \item \textit{Maintaining the full dynamics:} unlike dimensionality reduction methods, here the embedding has a higher dimension than the original system.
    
    \item \textit{Scalability:} the deep Koopman framework described in this paper can be trained on a higher dimensional system with more training data, without any modifications.
    
    \item \textit{Simplified linear analysis:} the deep Koopman framework facilitates the analysis of a nonlinear system by mapping it to a linear latent space where any linear analysis and control methodology can be applied to the system.
    
    \item \textit{Global interpretation:} unlike the classic method for dynamical analysis, this framework is not limited to local dynamics around equilibrium points or slow manifolds. Instead, it can characterize the global dynamics and make predictions about the future states of the system.
\end{itemize}

It is worth mentioning that adding physics-based constraints to the networks representing eigenfunctions and eigenvalues can result in a more physically interpretable analysis. For example, in \cite{Lusch.etal2018}, the authors designed the Koopman network to ensure purely oscillatory motion. Such analysis might provide insights into micro-macro relations in granular configurations. For instance, we expect certain properties of the granular configuration to be more relevant in achieving a desired wave response, and the above analysis can clarify this and provide guidelines to decide on certain aspects of parameter space. 

\newpage

\section{Acknowledgments}
This material is based upon work supported by the National Science Foundation under the DMREF program (award number: $2118810$). We thank the \href{https://www.uvm.edu/vacc}{Vermont Advanced Computing Center} for their computational resources.

\appendix*
\section{Details of Numerical Simulations}
\label{sec:appendix}
\begin{algorithm}[H]
\caption{Discrete Element Method}
\label{alg:DEM}
    \begin{algorithmic}
        \Require Initial positions ($\delta_{0, i}$, $u_i=0$) and velocities ($\dot{u}_{i}=0$), duration of the simulation ($T$), system's parameters (diameters, masses, stiffnesses, etc.) \\
        Initialize particle positions, velocities, accelerations, and forces
        \State $t \gets 0$
        \While{$t < T$}
            \For{all particles ($N$)}
                \State Compute total forces from the neighboring particles, walls, and external excitation;
                \State Integrate the equations of motion;
                \State Update accelerations, velocities, and positions;
            \EndFor
        \State $t \gets t + \Delta t$
        \EndWhile
    \end{algorithmic}
\end{algorithm}
\begin{algorithm}[H]
\caption{Velocity Verlet}
\label{alg:verlet}
\begin{algorithmic}
\Require Particle positions $r(t)$, velocities $\dot{r}(t)$ and accelerations $\ddot{r}(t)$ from the previous time step
    \State $r(t+\Delta t) \gets r(t) + [\dot{r}(t) + \frac{1}{2} \ddot{r}(t){\Delta t}] \Delta t = r(t) + \dot{r}(t) \Delta{t} + \frac{1}{2} \ddot{r}(t){\Delta t}^2$
    \State $\ddot{r}(t+\Delta t) \gets \frac{1}{m_i}F(r(t+\Delta t))$
    \State $\dot{r}(t+\Delta t) \gets \frac{1}{2} [\dot{r}(t-\frac{\Delta t}{2}) + \dot{r}(t+\frac{\Delta t}{2})] = \dot{r}(t) + \frac{\Delta t}{2} [\ddot{r}(t) + \ddot{r}(t+\Delta t)]$
\end{algorithmic}
\end{algorithm}

\bibliographystyle{unsrt}  
\bibliography{references}

\end{document}